\documentclass{article} 
\usepackage[T1]{fontenc}
\usepackage[final]{colm2026_conference}
\usepackage{booktabs, graphicx, subcaption}

\usepackage{amsmath, amssymb}
\usepackage{microtype}
\usepackage{hyperref}
\usepackage{url}
\usepackage{booktabs}
\usepackage{multirow}
\usepackage{rotating}
\usepackage{wrapfig}

\usepackage{lineno}

\definecolor{darkblue}{rgb}{0, 0, 0.5}
\hypersetup{colorlinks=true, citecolor=darkblue, linkcolor=darkblue, urlcolor=darkblue}

\title{Correcting test set contamination by spiking the training data}

\author{Johnny Tian-Zheng Wei*, Jerry Li*, Ameya Godbole, Robin Jia \\
University of Southern California \\
\\
\texttt{\{jtwei,lijc\}@usc.edu} \\
}

\begin{document}

\ifcolmsubmission
\linenumbers
\fi

\maketitle



\begin{abstract}
The literature on test set contamination largely focuses on detection, but the correction of contaminated test scores is underexplored. Our core proposal is to \textit{spike} the training data by intentionally contaminating some test examples at known rates. The spiked examples can then be used to calibrate predictors of model memorization which enable principled statistical correction of inflated test scores. To evaluate different correction estimators, we first present a simulation framework based on the Hubble models. Hubble models come in minimal pairs, where the perturbed model was deliberately contaminated with several test sets, while the standard model was not, serving as the counterfactual and correction target. We consider estimators that use information from a memorization predictor, correctness predictor, or both. In simulation, we establish basic statistical intuitions and show that estimators leveraging memorization and correctness information are better than naive estimation which makes no correction at all. We then instantiate several memorization and correctness predictors, and find that simple predictors such as Platt-scaled membership inference metrics provide good signal for correction. Finally, we examine the practical considerations of spiking. Simple memorization predictors need no more than 10 examples for calibration and often transfer from one dataset to another. Taken together, spiking is a promising solution for test set contamination.
\end{abstract}

\section{Introduction}

Measuring large language model (LLM) capabilities is difficult in part due to test set contamination \citep{magar-schwartz-2022-data, sainz-etal-2023-nlp}. The datasets used to train LLMs are vast and often include the test sets or original texts that the test sets were constructed from \citep{li-etal-2024-open-source, elazar2024whats}. As we look broadly to benchmarks for signals of general intelligence, disruptive economic potential, and safety \citep{ma2025benchmarking}, the accurate measurement of LLM capabilities is important. Test set contamination is actively studied \citep{conda-2024-data}, but most works focus on detecting contamination \citep[inter alia]{yang2023rethinkingbenchmarkcontaminationlanguage} or understanding contamination \citep[inter alia]{jiang2024investigatingdatacontaminationpretraining, schaeffer2026quantifyingeffecttestset}, rather than correcting inflated test scores. 


This work opens a line of inquiry on \textit{correcting} contaminated test scores.\footnote{All our code is available at: \url{https://github.com/Jeli04/spiking-tsc}.} We take the perspective that a model's test score is part true performance, part contamination. When a model answers a test item correctly, this raises two questions: whether the model memorized the answer due to contamination, and what it would have answered otherwise \citep{zhang_counterfactual_2023}. With a calibrated predictor of model memorization, we can isolate the true performance. Calibrating such a predictor requires examples where memorization status is randomized and known; therefore we propose \textit{spiking} the training data:  intentionally inserting some test examples at known rates to provide ground truth for calibration.

\begin{figure}[t]
    \centering
    \includegraphics[]{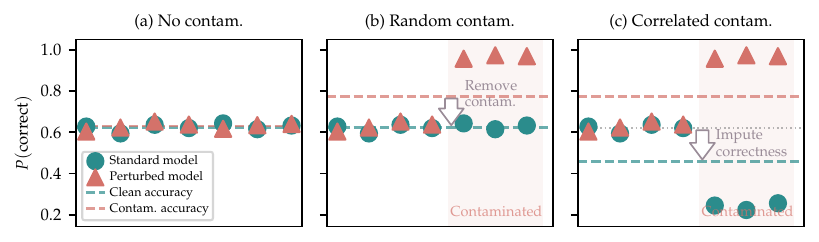}
    \caption{Illustration of correcting contaminated test scores. Each point represents a test example and its $P(\mathrm{correct})$ under the Hubble models. \textbf{(a)} When there is no contamination, the standard and perturbed model predict similarly. \textbf{(b)} The perturbed model memorizes contaminated items and the standard model serves as the ground truth. When test examples are randomly contaminated, removing contaminated items recovers the clean accuracy. \textbf{(c)} When contamination is correlated with difficulty (e.g. harder examples are more likely to be contaminated), removing contamination only partially recovers the clean accuracy. Fully recovering it additionally requires predicting the correctness of contaminated examples.}
    \label{fig:motivation}
\end{figure}

In \S\ref{sec:simulation}, we present a simulation framework to evaluate estimators for correcting contamination based on the Hubble models \citep{wei2026hubble}. Hubble models come in minimal pairs, where the perturbed models were intentionally contaminated with several test sets, but the standard model was not. In simulation, we repeatedly sample test sets with contaminated examples under the perturbed model, and the correction target is the standard model's clean, counterfactual accuracy on that same test set. We simulate settings where examples are either contaminated at random or correlated with example difficulty (see Figure \ref{fig:motivation}). Drawing inspiration from causal inference \citep{rubin_causal_2005}, we design several correction estimators relying on probabilistic predictors for memorization, correctness, or both. The simulation shows that different estimators are preferred in different scenarios, and we outline the basic statistical intuitions.

In \S\ref{sec:benchmarking}, we instantiate several memorization and correctness predictors and evaluate them in the simulation. To detect when the model is memorizing, we draw on the membership inference literature and evaluate the use of membership inference attacks as detectors \citep[MIAs;][]{DBLP:conf/sp/ShokriSSS17}. Once calibrated against known insertions, even simple MIAs provide useful signal to correct for contamination. To estimate correctness, we use a secondary LLM to provide signals on example difficulty \citep[similar to][]{dekoninck_2024_constat}. We evaluate both models we finetuned and pretrained LLMs, and the combined estimator using both sources of information performs best in many operating conditions. 

Finally, in \S\ref{sec:practical} we examine the practical considerations in spiking. Spiking requires items to be inserted into training, and we study how the size and distribution of the spiked set affect the estimators. Memorization predictors based on Platt scaling are remarkably sample efficient, and are well calibrated with just 10 examples. Memorization predictors calibrated on one dataset can generalize to others, and spiking Wikipedia passages is often sufficient to calibrate the memorization predictors. On the other hand, correctness predictors demand substantially more data. Taken together, the low cost of spiking and the robustness of memorization predictors point towards their promise in real-world use.

\section{Related work}

\textbf{Statistical perspectives on test set contamination are underexplored.} The study of test set contamination has yielded many insights on how contamination affects benchmark results \citep{magar-schwartz-2022-data, jiang2024investigatingdatacontaminationpretraining, schaeffer2026quantifyingeffecttestset}, and the community has developed a rich toolkit for detecting \citep{li-etal-2024-open-source, Li_Flanigan_2024, golchin_data_2025} and mitigating \citep{jacovi-etal-2023-stop, singh2024evaluationdatacontaminationllms} contamination. However, simply identifying which test examples are memorized is not enough for correction. Closest to our work, \cite{singh2024evaluationdatacontaminationllms} drops highly memorized examples and then re-estimates the clean accuracy. This reweighting is heuristic, and principled statistical estimation is possible by calibrating a memorization predictor.

\textbf{Calibration requires randomization.} To calibrate a detector, detector scores need to be matched against the likelihood that test examples are memorized \citep{platt1999probabilistic, guo_calibration_2017}. This requires a control population where memorization status is known, and several works approximate this control: \cite{golchin_data_2025} generate paraphrased versions of test examples and use them as clean, unseen data, and \cite{dekoninck_2024_constat} use reference models and reference benchmarks as proxies for uncontaminated performance. In both cases, these controls are synthetic and may introduce confounders \citep{mehrbakhsh-etal-2024-confounders}. Creating true controls requires randomization, and randomization is the basis of causal inference \citep{angrist_2009_mostly}. By inserting test examples at known rates, spiking introduces the randomization necessary for calibration and true statistical inference.

\textbf{Inserting randomization enables advanced auditing.} Randomization enables precise statistical inferences even when it arises naturally: \cite{oren2024proving} prove test sets were present in training data using the random ordering of the test examples, and \cite{zhu2025independence} prove one model is trained from another using the random training order. Our goal is to advance model metrology from the developer's perspective, assuming access to the training process \citep{saxon2024benchmarks}. Intentionally inserting randomization then enables a broader range of inferences \citep{steinke2023privacy, wei-etal-2024-proving}. Models like Apertus show that developers can be willing to insert canaries to study their model memorization and address stricter EU requirements \citep{apertus2025apertusdemocratizingopencompliant}, and \cite{bordt2026train} show that such insertions need not meaningfully harm model performance.

\section{Simulating contamination} \label{sec:simulation}

In this section, we formulate the correction problem and propose a simulation framework to evaluate correction estimators. The simulation is based on the Hubble models where we repeatedly sample test sets that are contaminated under the perturbed model, and for each contaminated example we also have the counterfactual and clean prediction from the standard model. The goal of estimators is to recover the accuracy of the standard model given the perturbed model, and we put forward several estimators using a memorization predictor, a correctness predictor, or both. By simulating synthetic predictors of varying quality, we first build intuition on when each estimator is appropriate.

\subsection{Estimators and predictors} \label{sec:estimators}

\textbf{Formulation.} In each simulated trial, we sample a test set of $n$ items. Items are contaminated with a contamination rate of $r_{\text{contam}}$ and each example is either clean or contaminated (duplicated at least once). Let $y_i \in \{0, 1\}$ denote whether the perturbed model was correct on example $i$, and let $y_i^* \in \{0, 1\}$ denote whether the standard model was correct. Our target is to recover the standard model accuracy $\mu^* = \frac{1}{n}\sum_{i=1}^n y_i^*$ using the perturbed model.

\textbf{Predictors.} Two types of predictors are necessary to correct for contamination. First, a \emph{memorization predictor} estimates $\hat{P}(\text{contam}|i)$, the probability that item $i$ is contaminated. Then, a \emph{correctness predictor} estimates $\hat{P}(\text{correct}|i) = \hat{P}(y^*_i=1|i)$, the probability that the (standard) model would have answered example $i$ correctly absent contamination. Memorization predictors are calibrated using known insertions, and correctness predictors are trained on clean items where $y_i^*$ is observed. Concrete instantiations of both predictors are benchmarked in \S\ref{sec:benchmarking}.

\textbf{Estimators.} We consider four estimators for $\mu^*$, drawing on ideas from causal inference.

The \emph{naive} estimator makes no correction:
\begin{equation}
    \hat{\mu}_{\text{naive}} = \frac{1}{n}\sum_{i=1}^n y_i
\end{equation}
and it overestimates the clean score in the presence of contamination.

The \emph{inverse propensity weighting} \cite[IPW;][]{rubin_causal_2005} estimator incorporates the memorization predictor and downweights items when they are memorized:
\begin{equation}
\hat{\mu}_{\text{ipw}} = \sum_{i=1}^n w_i \, y_i, \quad w_i = \frac{1 - \hat{P}(\text{contam} \mid i)}{\sum_{j=1}^n (1 - \hat{P}(\text{contam} \mid j))}
\end{equation}
This estimator ``drops'' the memorized items and depends on the discrimination of the memorization predictor, so we evaluate memorization predictors with AUROC. At best, the IPW estimator corrects to the average score of clean items in the test set, and if that does not match the true, clean score, then it will be biased (see Figure \ref{fig:motivation}, panel c).

The \emph{imputation} estimator replaces all outcomes with the correctness predictor:
\begin{equation}
    \hat{\mu}_{\text{imp}} = \frac{1}{n}\sum_{i=1}^n \hat{P}(\text{correct}|i)
\end{equation}
and this estimator depends entirely on the calibration of $\hat{P}(\text{correct}|i)$, so we evaluate correctness predictors by their absolute bias. The imputation estimator discards observed outcomes, and it is possible to achieve lower variance using a control variate estimator leveraging the clean observed outcomes \citep{mcbook}. However, we present a simpler estimator for the sake of exposition. 

The \emph{combined} estimator uses the memorization predictor to interpolate between the observed outcome and the prediction of the correctness predictor:
\begin{equation} \label{eq:combined}
\hat{\mu}_{\text{comb}} = \frac{1}{n}\sum_{i=1}^n \Big[\hat{P}(\text{contam}|i)\, \hat{P}(\text{correct}|i) + (1 - \hat{P}(\text{contam}|i))\, y_i\Big].
\end{equation}
This estimator is based on the law of total probability and uses the memorization predictor to route between the observed or the imputed outcome. For items that are likely clean ($\hat{P}(\text{contam}|i) \approx 0$), the observed outcome is trusted; and for items likely contaminated ($\hat{P}(\text{contam}|i) \approx 1$), the correctness predictor's prediction is substituted. This estimator is biased, but it corrects correlated contamination (see Figure~\ref{fig:motivation}, c) by imputing counterfactual outcomes for contaminated items rather than discarding them. Structurally, it resembles doubly robust methods from causal inference \citep{hlynsson2024tutorialdoublyrobustlearning}, which combine a propensity model with an outcome model. However, doubly robust methods require observing treatment assignment, whereas contamination status is unknown at test time.

\subsection{Data generation process} \label{sec:dgp}

\textbf{Hubble.} The Hubble model suite \citep{wei2026hubble} consists of pairs of standard and perturbed Llama-based LLMs. Standard models are trained on a standard English pretraining corpus and perturbed models are trained identically but with additional test examples from five benchmarks: WinoGrande, MMLU, HellaSwag, PIQA, and PopQA \citep{sakaguchi_winogrande_2021, hendrycks2021measuring, zellers-etal-2019-hellaswag, Bisk_Zellers_Lebras_Gao_Choi_2020, mallen-etal-2023-trust}. Our analyses focus on the 8B parameter models trained on 500B tokens as they perform best on these benchmarks. Both models were trained in the same way, and  for the perturbed model test examples were randomly assigned duplication rates in \{0, 1, 4, 16, 64, 256\}, where examples assigned a duplication rate of 0 were held out.

\textbf{Inserting canaries.} The core proposal of this paper is that model developers should spike their training data and intentionally insert some test examples at known rates. The spiked examples could then be used to calibrate probabilistic predictors for memorization. This assumes that we can observe some clean test examples, analogous to the positivity condition in causal inference (an assumption we revisit in \S\ref{sec:practical}). In our simulation, we reserve half of Hubble's test examples across each duplicate level as a spiked calibration set (\textasciitilde4000 items). The other half is the simulation set which is used to evaluate correction estimators.

\textbf{Simulation protocol.} From the simulation set (\textasciitilde4000 items), we sample test sets of $n = 500$ examples from each benchmark with a contamination rate of $r_{\text{contam}} = 0.3$. The clean items in each test set are drawn from held-out examples, and the contaminated items are drawn from duplicated examples. Varying levels of contamination strength can be simulated by sampling test examples that were more heavily duplicated. For each result we conduct 1{,}000 bootstrap simulations \citep{efron1994bootstrap}.

\textbf{Ground truth.} Our perspective on correcting for test set contamination is to remove performance gains due to contamination and isolate true performance. Since the standard model was not trained on inserted test examples, we assume it represents the true performance and take its predictions as the counterfactual ground truth. For each simulated test set, corrected estimates are compared to the standard model's accuracy on that test set, which includes the standard model predictions on clean examples as well.\footnote{This setup then has no spillover effects to correct for, as the standard model predictions are not affected by insertions of any test examples. Our implicit assumption is that spillover effects are part of generalization and unnecessary to correct for.} 

To achieve a low error, correction estimators would have to contend with both contamination and training stochasticity as the perturbed and standard models occasionally disagree, even on clean examples. To eliminate this source of error, the observed outcomes $y_i$ for clean items are based on the standard model and the observed outcomes $y_i$ on contaminated items are based on the perturbed model. In practice, developers only have a single contaminated model, and contamination is the only term that requires correction. 

\subsection{Phase diagrams} \label{sec:phase}

\begin{figure}[t]
    \centering
    \includegraphics[]{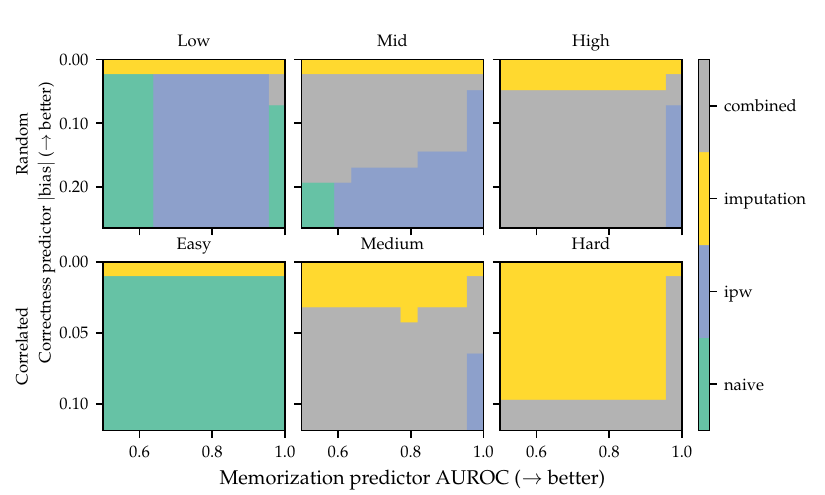}
    \caption{Phase diagrams showing winning estimators by lowest RMSE, under varying memorization and correctness predictor quality. Estimators are described in \S\ref{sec:estimators} and this simulation uses MMLU data and synthetic predictors  (details in \S\ref{sec:phase}). \textbf{(Top row)} Random contamination at increasing contamination strength (low = 1$\times$ duplicates, mid = 16$\times$, high = 64/256$\times$). As the strength increases, the combined estimator's optimal region grows. \textbf{(Bottom row)} Correlated contamination at mid and high contamination, where contaminated examples are easy, medium, or hard. When the contaminated items are hard, IPW fails because removing contaminated items removes difficult examples and introduces selection bias. Only the combined or imputation estimator can recover the clean score.}
    \label{fig:phase_diagrams}
\end{figure}

\textbf{Contamination regimes.} In each trial, we sample $n=500$ test examples with contamination rate $r_{\text{contam}} = 0.3$. Contaminated examples are sampled under two regimes:

In \emph{random contamination} (Figure \ref{fig:motivation}, panel b), items are contaminated independently of their difficulty, with increasing memorization strength at low (1 duplicate), medium (16), or high (64 or 256). Clean items remain representative of the entire test set, so removing contaminated items suffices to recover the clean score. 

In \emph{correlated contamination} (Figure \ref{fig:motivation}, panel c), contamination correlates with difficulty. Highly memorized examples (64 and 256 duplicates) are binned into easy, medium, and hard based on the standard model's confidence, and contaminated examples are drawn from these bins. Here, clean items are no longer representative and recovering the clean score additionally requires imputing counterfactual outcomes for contaminated items.

\textbf{Synthetic predictors.} To illustrate how estimators depend on predictor quality, we simulate synthetic predictors of varying levels of quality. Both predictors convert ground truth labels to continuous prediction scores by drawing from two beta distributions. For the positive class samples are drawn from one beta distribution, and for the negative class samples are drawn from the other. For the memorization predictor, we set the parameters of the class-conditional Beta distributions to achieve a target AUROC. For the correctness predictor, we set the parameters to achieve a target bias, and it interpolates between the ground truth (bias=0) and the uniform distribution. In both cases, a shared concentration parameter governs the score variance around the class means.

\textbf{Results.} Figure \ref{fig:phase_diagrams} presents the simulation results with synthetic predictors for MMLU. The phase diagrams of winning estimators provide us basic statistical intuitions:

\textit{Under stronger contamination, the combined estimator is preferred.} Looking at the random contamination regime, as the strength of contamination increases, the region of the combined estimator grows. Conversely, the region of the naive estimator shrinks when there is stronger contamination. Under random contamination, using naive is rarely optimal, and incorporating either memorization or correctness information will provide better estimates.

\textit{Asymmetry between IPW and imputation.} Under random contamination, the optimal region of IPW (drops memorized examples) is larger than that of the imputation estimator (imputes labels for all examples). There are two reasons: first, the imputation estimator applies the correctness predictor to all $n$ items, including those that are clean and did not need correction. Each unnecessary imputation contributes to the final estimate and accumulates errors, while IPW uses the observed outcomes for items it retains. Second, the correctness predictor contains random noise by construction, so it has attenuation bias towards the uniform distribution.

\textit{Correlated contamination presents challenges for estimators.} As illustrated in Figure \ref{fig:motivation}, dropping memorized items is not sufficient when contamination correlates with example difficulty. From the easy to hard regime, there is a large swing in which estimators are optimal. In the easy regime, the naive estimator is especially advantaged as contamination does not change the outcome on items the model would have answered correctly anyway. In the hard regime, where difficult items are contaminated, contamination flips those examples from incorrect to correct, and simply dropping them introduces selection bias. Recovering the clean score here requires imputing counterfactual outcomes, and the imputation estimator has a much larger feasible region.

\section{Benchmarking predictors} \label{sec:benchmarking}

The previous section used synthetic predictors to understand how estimators perform under varying predictor quality. In this section, we instantiate concrete memorization and correctness predictors and deploy them in the simulation. Our goal here is not necessarily to train the best predictors, but to understand whether baseline methods provide enough signal for correction. Any calibration or training uses the calibration split described in \S\ref{sec:dgp}.

\subsection{Memorization predictors}

\textbf{Membership inference attacks.} To predict whether an example is memorized, we  draw on the rich literature of membership inference attacks (MIAs). MIAs are designed to detect whether a given example was present in a model's training data \citep{DBLP:conf/sp/ShokriSSS17}, which makes them suitable memorization predictors. We evaluate four simple attacks derived from token-level statistics in a single forward pass across both question and answer tokens. Each attack produces a raw score that we calibrate to a probability using Platt scaling on the calibration examples \citep{platt1999probabilistic}. They are listed below:

\begin{itemize}
    \item \textbf{LOSS.} This is the simplest baseline and is just the sequence log-likelihood averaged over the number of tokens \citep{10.3233/JCS-191362}.
    \item \textbf{Min-K\%.} Averages the minimum-$k\%$ of token log-probabilities in the sequence, with each token ranked by its own raw log-probability \citep{shi2024detecting}.
    \item \textbf{Min-K\%++.} Averages the minimum-$k\%$ of token log-probabilities in the sequence, where each token's log probability is normalized by the variance of the model's conditional distribution at that position \citep{zhang2025mink}.
    \item \textbf{zlib.} Mean log-likelihood normalized by the zlib-compressed byte length of the input text \citep{274574}. This metric uses zlib as a reference  model to correct for the intrinsic compressibility of a sequence. 
    \item \textbf{Reference (Oracle).} Log-likelihood ratio between the target model and a clean reference model \citep{274574}. We use the standard model as the reference model to serve as an oracle upper bound.
\end{itemize}

\begin{table*}[t]
\centering
\small
\setlength{\tabcolsep}{3.5pt}
\begin{tabular}{lcccccccccccc}
\toprule
\textbf{Predictor} & \multicolumn{4}{c}{\textbf{WinoGrande}} & \multicolumn{4}{c}{\textbf{MMLU}} & \multicolumn{4}{c}{\textbf{PopQA}} \\
\cmidrule(lr){2-5} \cmidrule(lr){6-9} \cmidrule(lr){10-13}
 & \textbf{Low} & \textbf{Med} & \textbf{High} & \textbf{All} & \textbf{Low} & \textbf{Med} & \textbf{High} & \textbf{All} & \textbf{Low} & \textbf{Med} & \textbf{High} & \textbf{All} \\
\midrule
Reference* & \textbf{0.555} & \textbf{0.731} & \textbf{0.995} & \textbf{0.697} & 0.529 & 0.741 & 0.974 & 0.691 & \textbf{0.560} & \textbf{0.728} & 0.967 & \textbf{0.694} \\
LOSS & 0.534 & 0.689 & \textbf{0.995} & 0.666 & 0.532 & 0.720 & \textbf{1.000} & 0.683 & 0.553 & 0.720 & \textbf{0.997} & 0.690 \\
Min-K\%++ & 0.534 & 0.684 & 0.987 & 0.663 & \textbf{0.563} & \textbf{0.789} & \textbf{1.000} & \textbf{0.731} & 0.551 & 0.724 & 0.995 & 0.691 \\
Min-K\% & 0.522 & 0.661 & 0.982 & 0.646 & 0.549 & 0.769 & \textbf{1.000} & 0.715 & 0.555 & 0.724 & \textbf{0.997} & 0.693 \\
zlib & 0.524 & 0.677 & 0.992 & 0.656 & 0.539 & 0.679 & 0.995 & 0.663 & 0.546 & 0.700 & 0.991 & 0.677 \\
\bottomrule
\end{tabular}
\caption{Memorization predictor AUROC results for discriminating contamination. Results are shown under the random contamination regime (low, medium, high), with the aggregate binary AUROC shown in the "all" column for each benchmark. At low contamination, memorization predictors do not discriminate well but are near perfect for highly contaminated examples. Reference* denotes an oracle method, which uses the standard Hubble model.}
\label{tab:mem_results}
\end{table*}

\textbf{Results.} Table \ref{tab:mem_results} reports evaluation results on the simulation set. Memorization predictors are trained on test examples from the calibration set. Stronger contamination is easier to detect. All methods perform near chance when distinguishing between clean examples and lightly contaminated examples, consistent with the findings in \cite{duan2024membership}. However, all methods detect heavy contamination reliably. Even the LOSS metric after Platt scaling works well. Since contamination in Hubble is defined by exact duplication, likelihood-based signals are a strong indicator of memorization.

\begin{table*}[t]
\centering
\small
\setlength{\tabcolsep}{3.5pt}
\begin{tabular}{lcccccccccccc}
\toprule
\textbf{Predictor} & \multicolumn{4}{c}{\textbf{WinoGrande }} & \multicolumn{4}{c}{\textbf{MMLU}} & \multicolumn{4}{c}{\textbf{PopQA}} \\
\cmidrule(lr){2-5} \cmidrule(lr){6-9} \cmidrule(lr){10-13}
 & \textbf{Easy} & \textbf{Med} & \textbf{Hard} & \textbf{All}
 & \textbf{Easy} & \textbf{Med} & \textbf{Hard} & \textbf{All}
 & \textbf{Easy} & \textbf{Med} & \textbf{Hard} & \textbf{All} \\
\midrule
Llama-3.1
& \textbf{0.019} & \textbf{0.017} & \textbf{0.097} & 0.032
& \textbf{0.137} & 0.047 & \textbf{0.018} & 0.068
& \textbf{0.158} & 0.107 & \textbf{0.002} & 0.089 \\
RoBERTa
& 0.081 & 0.049 & 0.164 & \textbf{0.010}
& 0.203 & \textbf{0.004} & 0.216 & \textbf{0.003}
& 0.236 & \textbf{0.034} & 0.114 & 0.029 \\
Pythia 6.9b
& 0.043 & 0.052 & 0.129 & 0.011
& 0.220 & 0.016 & 0.201 & 0.011
& 0.167 & 0.074 & 0.087 & 0.002 \\
Qwen3 8b
& 0.036 & 0.046 & 0.126 & 0.014
& 0.163 & 0.013 & 0.150 & 0.009
& 0.121 & 0.063 & 0.058 & \textbf{0.000} \\
\bottomrule
\end{tabular}
\caption{Absolute bias of correctness predictors under correlated contamination, broken down by difficulty bin, with averages shown in the “all” column. The all column is representative of predictor bias in the random contamination regime. Lower values indicate less biased predictions. RoBERTa is a finetuned method, while Llama 3.1, Pythia 6.9B, and Qwen 3 8B are pretrained and use Platt scaling. Correctness predictors generally have low bias over the entire dataset.}
\label{tab:corr_results}
\end{table*}

\subsection{Correctness predictors}

Correctness predictors estimate the counterfactual outcome, predicting whether the model would have answered correctly absent contamination. In part, this requires estimating example difficulty, which is a well-studied problem, and statistical frameworks use multiple responses to a test item to isolate the difficulty of that item \citep{pmlr-v162-ethayarajh22a, DBLP:conf/icml/TruongTL0K25}. Similarly, we study using another LLM to estimate the difficulty of the examples. Unlike the memorization predictors, correctness predictors are trained or calibrated on only the clean samples from the calibration split, which are standard model predictions (see \S\ref{sec:dgp}). 

\textbf{RoBERTa for sequence classification.} We fine-tune RoBERTa \citep{liu2019robertarobustlyoptimizedbert} with a classification head to predict the correctness of the Hubble model’s response. Training is performed on the calibration split, using only the test questions as input. We use a learning rate of $5 \times 10^{-6}$ and a batch size of 32, while freezing the first five layers. This configuration was chosen based on a hyperparameter search using test performance on the simulation set.

\textbf{LLM with Platt scaling.} Instead of finetuning a correctness predictor, we can pair the Hubble model against another pretrained LLM to predict its correctness. We take the confidence of the correct answer from the paired LLMs and use Platt scaling to match it to the likelihood of Hubble's correctness. For the paired models, we evaluate Llama 3.1, Pythia 6.9B, and Qwen3 8B \citep{grattafiori2024llama3herdmodels,biderman2023pythia,qwen3technicalreport}.

\textbf{Results.} Table~\ref{tab:corr_results} reports correctness predictor bias by benchmark and difficulty level. Llama 3.1 generally shows lower absolute bias than RoBERTa on WinoGrande, whereas MMLU and PopQA are more mixed. Bias increases in the hard contamination setting, where the perturbed model diverges most from the standard model. Notably, the average bias remains low across benchmarks, placing all predictors in the region of the phase diagrams where correction is beneficial.

\subsection{Correcting estimates with predictors}

\begin{table*}[t]
\centering
\small
\setlength{\tabcolsep}{8pt}
\begin{tabular}{lcccccc}
\toprule
\multirow{2}{*}{\textbf{Estimator}} & \multicolumn{3}{c}{\textbf{Random}} & \multicolumn{3}{c}{\textbf{Correlated (high dose)}} \\
\cmidrule(lr){2-4} \cmidrule(lr){5-7}
& \textbf{Low} & \textbf{Mid} & \textbf{High} & \textbf{Easy} & \textbf{Medium} & \textbf{Hard} \\
\midrule
Naive                             & 1.8 & 6.4 & 13.1 & \textbf{0.5} & 10.3 & 29.2 \\
EPG (Min-K\%++)                   & 7.4 & 4.3 & 2.1 & 14.8 & 3.6 & 15.3 \\
\midrule
IPW (Min-K\%++)                   & \textbf{1.7} & 4.5 & 6.4 & 7.5 & 4.1 & 22.3 \\
Imputation (Qwen3 8B)             & 3.5 & 4.3 & 4.1 & 15.5 & 6.1 & \textbf{10.5} \\
Combined (Min-K\%++ \& Qwen3 8B)  & 2.3 & \textbf{1.4} & \textbf{1.8} & 11.2 & \textbf{2.0} & 15.4 \\
\bottomrule
\end{tabular}
\caption{Simulated estimator performance on MMLU under random and correlated contamination regimes. Values are RMSE against the ground-truth standard model accuracy over 1,000 bootstrap replicates. In other words, naive overpredicts the clean test set accuracy under random, high contamination by 13.1 points on average. Estimators using a correctness or memorization predictor reliably adjust for contamination, with the combined estimator using both predictors winning in most settings.}
\label{tab:estimator_random_correlated}
\end{table*}

\textbf{Results.} We evaluate the estimators with the memorization and correctness predictors in the simulation and report the best estimators (which use Min-K\%++, Qwen3 8b for the predictors). The results on MMLU are given in Table \ref{tab:estimator_random_correlated} and the other benchmarks are reported in Appendix~\ref{app:adjustment_simulation}.

\textit{There are substantial gains over the naive estimator.} In almost every setting, the correction estimators match or substantially improve on the naive estimator, with the largest gains under high contamination. Even IPW is quite effective, which only requires a calibrated MIA. The combined estimator is competitive across settings while avoiding the worst case failures of either single-source estimator.

\textit{Low contamination is difficult to correct, but also introduces little error.} At the low end, most estimators only match or perform worse than the naive estimator. As seen in the simulation, naive has a larger optimal region when the contamination is lower, and correcting for smaller differences is more difficult and requires more precision. While estimators perform worse here, the cost of not correcting is also low, as light contamination does not inflate performance much \citep[][make similar points]{bordt2025how, wei2026hubble}.

\textit{Correlated contamination presents mixed results}. Under correlated contamination, estimator performance depends heavily on which examples are contaminated. When easy examples are contaminated, the naive estimator is nearly optimal, consistent with the phase diagrams in Figure \ref{fig:phase_diagrams}. When hard examples are contaminated, score inflation is severe and the naive estimator incurs the largest error. IPW is also insufficient in this setting, as dropping contaminated hard items introduces selection bias. The combined and imputation estimators provide substantial gains by imputing counterfactual outcomes for contaminated items rather than discarding them, though residual error remains high.

\textit{Heuristic methods do not provide consistent correction.} The reweighting method in \citet{singh2024evaluationdatacontaminationllms} chooses the threshold (over Min-K\%++ scores) that jointly maximizes the expected performance gain (EPG; details in Appendix~\ref{app:epg}). This method does not require spiking but chooses the decision boundary for clean and contaminated examples heuristically. The gain of EPG over naive estimation is inconsistent across random contamination, while spiking-based IPW leads to consistent improvement.

\section{Practical considerations} \label{sec:practical}

In the previous section, half of the contaminated test examples in Hubble were held out as the spiked calibration set, and experiments used all of it for calibration (approximately 4{,}000 items per dataset). Spiking requires inserting a clean set of items into training, which may be difficult to meet in practice. This section examines the practical considerations of spiking, and we study how varying the size and distribution of the spiked set affect the estimators.

\begin{figure}[t]
    \centering
    \includegraphics[width=\columnwidth]{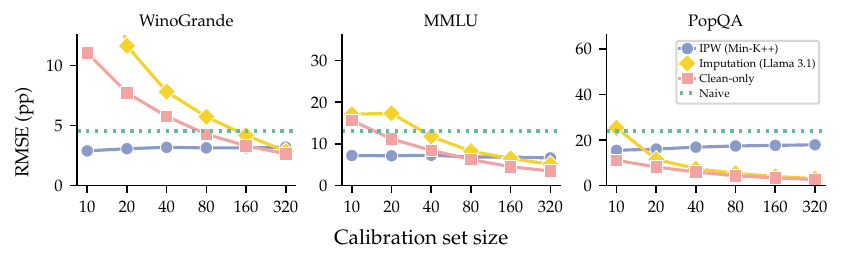}
    \caption{RMSE of IPW (Min-K\%++) and imputation (Llama 3.1) estimators while varying the size of the spiked calibration set. Simulations use high random contamination and are comparable to those of the same setting in Table \ref{tab:estimator_random_correlated}. The clean-only estimator uses the calibration items as a new test set and serves as a baseline, while the naive estimator applies no correction. Since items can be spiked at different rates, the calibration set for the memorization predictor is balanced and the IPW estimator is extremely sample efficient.}
    \label{fig:practical_efficiency}
\end{figure}

\textbf{Sample efficiency.} Obtaining more test samples from a benchmark can entail significant effort or resources \citep{pmlr-v97-recht19a}, and an important practical question is how many spiked examples are actually needed. One natural baseline is then the clean-only estimator, which uses the spiked items purely as a new, clean test set rather than for calibration. Figure \ref{fig:practical_efficiency} shows the sample efficiency of different estimators under random, high contamination. The IPW estimator calibrated with Min-K\%++ needs only 10 examples to be effectively calibrated, and this is due to the simplicity of Platt scaling which fits only two parameters. Examples are also spiked across contamination levels to provide a balanced calibration set. The correctness predictor, by contrast, requires several hundred examples before the imputation estimator matches or surpasses IPW. Although IPW is not as strong as the combined estimator in Table \ref{tab:estimator_random_correlated}, its low calibration cost makes it a strong default option.

\begin{table}[t]
\centering
\small
\setlength{\tabcolsep}{3.5pt}
\begin{tabular}{lcccccc}
\toprule
& \multicolumn{3}{c}{\textbf{Mid}} & \multicolumn{3}{c}{\textbf{High}} \\
\cmidrule(lr){2-4} \cmidrule(lr){5-7}
\textbf{Source} & \textbf{WinoGrande} & \textbf{MMLU} & \textbf{PopQA} & \textbf{WinoGrande} & \textbf{MMLU} & \textbf{PopQA} \\
\midrule
Naive       & 3.0 & 6.4 & 11.6 & 4.6 & 13.1 & 24.0 \\
\midrule
WinoGrande          & 1.9 & 5.8 & \textbf{2.4} & 2.0 & 5.2 & \textbf{3.1} \\
MMLU        & 2.3 & \textbf{5.6} & 4.3 & 2.4 & \textbf{4.7} & 5.8 \\
PopQA       & 2.2 & 5.8 & 5.0 & 2.6 & 6.7 & 8.6 \\
Wikipedia   & \textbf{1.3} & 6.9 & 3.9 & \textbf{1.4} & 5.5 & 5.8 \\
\bottomrule
\end{tabular}
\caption{RMSE of the IPW estimator when the memorization predictor (Min-K\%) is transferred across datasets. The memorization predictor is calibrated and evaluated on different benchmarks. These simulations are comparable to those of the same settings in Table \ref{tab:estimator_random_correlated}. At high contamination, transferred predictors almost always improve over the naive estimator, and predictors calibrated on Wikipedia texts closely match dataset-specific calibration.}
\label{tab:practical_transfer}
\end{table}

\textbf{Distribution shift.} Once a model is trained, the spiked set is fixed but developers may want to correct for contamination in new benchmarks. In that scenario, a memorization predictor calibrated on one set would be applied to another. Table~\ref{tab:practical_transfer} presents results transferring IPW predictors (using Min-K\%) from one dataset to another. Transferring Min-K\% across datasets is generally effective, and transferred performance is comparable to in-domain calibration. Min-K\%++ is less transferable due to dataset-specific normalization (results in Appendix \ref{app:transfer}). At high contamination, transferred predictors almost always improve over the naive estimator, and predictors calibrated on Wikipedia texts closely matches dataset-specific calibration. This suggests that memorization predictors trained on a general corpus could still be effective for new test sets, reducing the burden on developers to spike every test set individually.

\section{Conclusion}

This work opens a line of inquiry on correcting contaminated benchmark scores. With the simulation framework based on the Hubble models, we can evaluate estimators against a ground truth. The simulation shows that correction estimators substantially reduce error relative to the naive baseline, and even simple Platt-scaled memorization and correctness predictors with two trained parameters are effective. Overall, spiking is a promising solution that deserves future study. Due to how the Hubble models are contaminated, our simulation is limited to studying contamination through exact duplicates. In practice, test sets are also contaminated through paraphrases or near-duplicates \citep{jiang2024investigatingdatacontaminationpretraining}, and this simulation can be extended with paired models trained with other forms of contamination. 

Randomization is the foundation of advanced auditing. Spiking inserts the randomization necessary to enable a wider range of inferences. Besides correcting for test set contamination, it is the basis for other complex inferences as well \citep[e.g. canaries are used in][to estimate differential privacy parameters]{steinke2023privacy}. As a training-side intervention, spiking requires cooperation from model developers. We believe that spiking will play a key role in technical governance \citep{reuel2025open}. Black-box access to a model is insufficient for rigorous audits \citep{casper_black_2024}, but releasing spiked data is a compromise which allows auditors and researchers to run a wide range of analyses on the models. Standards-setting bodies have begun to formalize practices for automated benchmark evaluation \citep{nist2026ai800-2}, and we call on the community to develop a literature around spiking-based evaluation. If the evidence shows that the benefits outweigh the costs, spiking may eventually become best practice and see widespread use.

\textbf{Disclosure: the authors used LLMs for coding, plotting, ideation.}

\section*{Acknowledgments}
We thank Kyle Ng and anonymous reviewers for their feedback on this work. 

This research is based upon work supported in part by the Office of the Director of National Intelligence (ODNI), Intelligence Advanced Research Projects Activity (IARPA), via 56000026C0020. The views and conclusions contained herein are those of the authors and should not be interpreted as necessarily representing the official policies, either expressed or implied, of ODNI, IARPA, or the U.S. Government. The U.S. Government is authorized to reproduce and distribute reprints for governmental purposes notwithstanding any copyright annotation therein.
This work was supported in part by the National Science Foundation under Grant No. IIS-2403436. Any opinions, findings, and conclusions or recommendations expressed in this material are those of the author(s) and do not necessarily reflect the views of the National Science Foundation.
This work was also supported in part by a gift from the USC-Amazon Center on Secure and Trusted Machine Learning. 

\bibliography{colm2026_conference}
\bibliographystyle{colm2026_conference}

\newpage
\appendix

\section{Non-spiking Baseline: EPG}
\label{app:epg}

\citet{singh2024evaluationdatacontaminationllms} propose a heuristic rule to threshold the scores from a contamination detection method and separate the `clean and contaminated examples. The heuristic is based on the assumption that test set contamination will artificially inflate the performance of a model on the full test set. For a contamination detection method, the method selects the optimal threshold that maximizes:

\begin{equation}
\text{z-score}(t) = \frac{\text{EPG}(t)}{\text{err}(t)}
\end{equation}

$\text{EPG}(t)$ is the ``estimated performance gain" from the (presumably contaminated) examples with detector scores above the threshold $t$. The denominator $\text{err}(t)$ is the standard error for the clean subset size at threshold $t$ ($\text{err}(t) = \sigma / \sqrt{N(t)}$ where $\sigma$ is the standard deviation on the full benchmark and $N(t)$ is the size of the test set marked ``clean" at the threshold $t$). Intuitively, the threshold that maximizes the z-score drops the examples that contribute most to the inflated accuracy while preferring a threshold that drops the fewest examples. \citet{singh2024evaluationdatacontaminationllms} only study contamination detectors based on n-gram overlap metrics. We extend their technique to use membership inference predictors as contamination detectors as n-gram overlap statistics rely on access to the pre-training corpus.

Since EPG does not use Platt scaling, it often overestimates the amount of contamination in the test sets. In our experiments across all datasets and regimes, EPG estimates $\sim$60\% of examples as contaminated (despite the underlying contamination rate of 30\% in our simulated benchmarks). Notable exceptions are MMLU high-dose (random and all correlated) regimes where the optimal threshold marks 4.7\% and 32\% examples as contaminated respectively. This allows it to perform competitively with the Platt-scaled IPW (Min-K\%++) on MMLU. However, from Table~\ref{tab:epg_all}, we can see the lack of consistency across benchmarks establishing the necessity of spiking.

\clearpage
\section{Additional results}
\subsection{Phase diagrams}
\begin{figure}[h]
    \centering
    \includegraphics[]{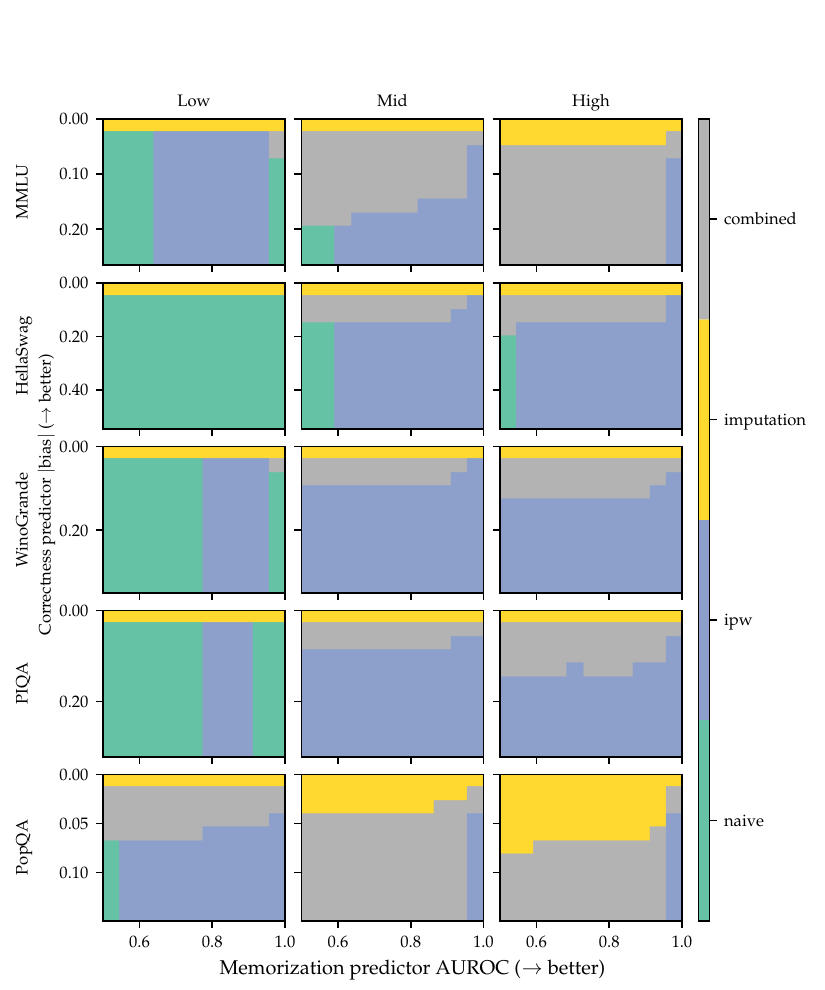}
    \caption{Phase diagrams showing winning estimators by lowest RMSE, under varying memorization and correctness predictor quality, for all datasets. Estimators are described in \S\ref{sec:estimators} and use synthetic predictors (details in \S\ref{sec:phase}). Random contamination at increasing contamination strength (low = 1$\times$ duplicates, mid = 16$\times$, high = 64/256$\times$). As contamination strength increases, the combined estimator's optimal region grows.}
    \label{fig:appendix_random}
\end{figure}

\clearpage
\begin{figure}[h]
    \centering
    \includegraphics[]{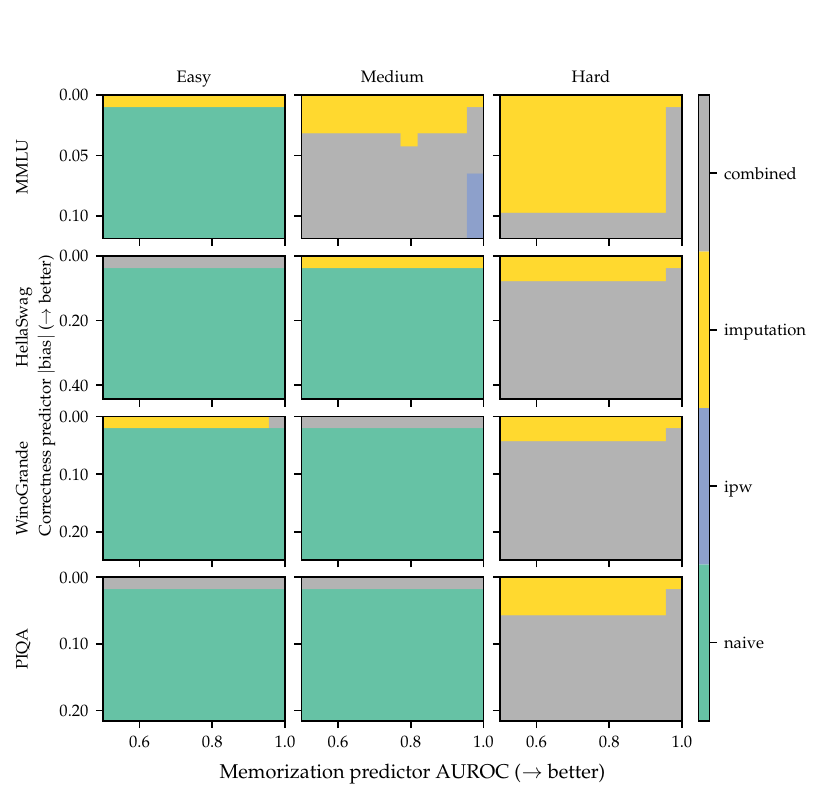}
    \caption{Phase diagrams showing winning estimators by lowest RMSE, under varying memorization and correctness predictor quality, for all datasets. Estimators are described in \S\ref{sec:estimators} and use synthetic predictors (details in \S\ref{sec:phase}). Correlated contamination at mid and high contamination, where contaminated examples are easy, medium, or hard. When the contaminated items are hard, IPW fails because removing contaminated items removes difficult examples and introduces selection bias. Only the combined or imputation estimator can recover the clean score.}
    \label{fig:appendix_correlated}
\end{figure}
\clearpage

\subsection{Benchmarking predictors}
\begin{table*}[h]
\centering
\small
\setlength{\tabcolsep}{8pt}
\renewcommand{\arraystretch}{1.2}
\begin{tabular}{lcccc}
\toprule
\textbf{Predictor} & \textbf{Easy} & \textbf{Med} & \textbf{Hard} & \textbf{All} \\
\midrule
\multicolumn{5}{c}{\textbf{WinoGrande}} \\
\midrule
Llama 3.1 & \textbf{0.0187} & \textbf{0.0169} & \textbf{0.0966} & 0.0323 \\
RoBERTa & 0.0814 & 0.0488 & 0.1642 & \textbf{0.0103} \\
Pythia 6.9B & 0.0425 & 0.0521 & 0.1293 & 0.0107 \\
Qwen3 8B & 0.0364 & 0.0460 & 0.1256 & 0.0136 \\
\midrule
\multicolumn{5}{c}{\textbf{MMLU}} \\
\midrule
Llama 3.1 & \textbf{0.1370} & 0.0474 & \textbf{0.0182} & 0.0676 \\
RoBERTa & 0.2032 & \textbf{0.0041} & 0.2156 & \textbf{0.0027} \\
Pythia 6.9B & 0.2199 & 0.0157 & 0.2014 & 0.0115 \\
Qwen3 8B & 0.1628 & 0.0133 & 0.1497 & 0.0089 \\
\midrule
\multicolumn{5}{c}{\textbf{PIQA}} \\
\midrule
Llama 3.1 & 0.0595 & \textbf{0.0254} & 0.0727 & 0.0357 \\
RoBERTa & 0.0598 & 0.0774 & 0.1920 & 0.0182 \\
Pythia 6.9B & \textbf{0.0137} & 0.0674 & \textbf{0.0720} & \textbf{0.0029} \\
Qwen3 8B & 0.0138 & 0.0712 & 0.0736 & 0.0037 \\
\midrule
\multicolumn{5}{c}{\textbf{HellaSwag}} \\
\midrule
Llama 3.1 & 0.0762 & \textbf{0.0159} & 0.1418 & 0.0780 \\
RoBERTa & 0.0932 & 0.0921 & 0.2955 & 0.0367 \\
Pythia 6.9B & \textbf{0.0209} & 0.0565 & 0.1027 & 0.0084 \\
Qwen3 8B & 0.0250 & 0.0522 & \textbf{0.0983} & \textbf{0.0070} \\
\midrule
\multicolumn{5}{c}{\textbf{PopQA}} \\
\midrule
Llama 3.1 & 0.1581 & 0.1070 & \textbf{0.0023} & 0.0893 \\
RoBERTa & 0.2355 & \textbf{0.0338} & 0.1143 & 0.0294 \\
Pythia 6.9B & 0.1666 & 0.0740 & 0.0868 & 0.0021 \\
Qwen3 8B & \textbf{0.1206} & 0.0629 & 0.0579 & \textbf{0.0001} \\
\bottomrule
\end{tabular}
\caption{Absolute bias of correctness predictors under correlated contamination, for all datasets, broken down by difficulty bin, with averages shown in the ``all'' column. The all column is representative of predictor bias in the random contamination regime. Lower values indicate less biased predictions. RoBERTa is a finetuned method, while Llama 3.1, Pythia 6.9B, and Qwen 3 8B are pretrained and use Platt scaling. Correctness predictors generally have low bias over the entire dataset.}
\label{tab:corr_abs_bias}
\end{table*}

\begin{table*}[t]
\centering
\small

\begin{tabular}{lcccc}
\toprule
\multicolumn{5}{c}{\textbf{WinoGrande}} \\
\midrule
\textbf{Predictor} & \textbf{Low} & \textbf{Med} & \textbf{High} & \textbf{All} \\
\midrule
Reference* & \textbf{0.525} & \textbf{0.702} & 0.995 & \textbf{0.671} \\
LOSS       & 0.521 & 0.686 & 0.997 & 0.661 \\
Min-K\%++  & 0.516 & 0.701 & 0.993 & 0.666 \\
Min-K\%    & 0.512 & 0.685 & 0.991 & 0.656 \\
zlib       & 0.521 & 0.686 & \textbf{0.997} & 0.660 \\
\bottomrule
\end{tabular}

\vspace{0.9em}

\begin{tabular}{lcccc}
\toprule
\multicolumn{5}{c}{\textbf{MMLU}} \\
\midrule
\textbf{Predictor} & \textbf{Low} & \textbf{Med} & \textbf{High} & \textbf{All} \\
\midrule
Reference* & 0.529 & 0.741 & 0.974 & 0.691 \\
LOSS       & 0.532 & 0.720 & 1.000 & 0.683 \\
Min-K\%++  & \textbf{0.563} & \textbf{0.789} & 1.000 & \textbf{0.731} \\
Min-K\%    & 0.549 & 0.769 & \textbf{1.000} & 0.715 \\
zlib       & 0.539 & 0.679 & 0.995 & 0.663 \\
\bottomrule
\end{tabular}

\vspace{0.9em}

\begin{tabular}{lcccc}
\toprule
\multicolumn{5}{c}{\textbf{PIQA}} \\
\midrule
\textbf{Predictor} & \textbf{Low} & \textbf{Med} & \textbf{High} & \textbf{All} \\
\midrule
Reference* & \textbf{0.585} & \textbf{0.783} & 1.000 & \textbf{0.735} \\
LOSS       & 0.560 & 0.743 & \textbf{1.000} & 0.706 \\
Min-K\%++  & 0.566 & 0.757 & 0.997 & 0.714 \\
Min-K\%    & 0.543 & 0.740 & \textbf{1.000} & 0.698 \\
zlib       & 0.527 & 0.664 & 0.997 & 0.651 \\
\bottomrule
\end{tabular}

\vspace{0.9em}

\begin{tabular}{lcccc}
\toprule
\multicolumn{5}{c}{\textbf{PopQA}} \\
\midrule
\textbf{Predictor} & \textbf{Low} & \textbf{Med} & \textbf{High} & \textbf{All} \\
\midrule
Reference* & \textbf{0.560} & \textbf{0.728} & 0.967 & \textbf{0.694} \\
LOSS       & 0.553 & 0.720 & \textbf{0.997} & 0.690 \\
Min-K\%++  & 0.551 & 0.724 & 0.995 & 0.691 \\
Min-K\%    & 0.555 & 0.724 & 0.997 & 0.693 \\
zlib       & 0.546 & 0.700 & 0.991 & 0.677 \\
\bottomrule
\end{tabular}

\vspace{0.9em}

\begin{tabular}{lcccc}
\toprule
\multicolumn{5}{c}{\textbf{HellaSwag}} \\
\midrule
\textbf{Predictor} & \textbf{Low} & \textbf{Med} & \textbf{High} & \textbf{All} \\
\midrule
Reference* & \textbf{0.599} & \textbf{0.825} & 1.000 & \textbf{0.763} \\
LOSS       & 0.526 & 0.721 & \textbf{1.000} & 0.681 \\
Min-K\%++  & 0.550 & 0.764 & 1.000 & 0.713 \\
Min-K\%    & 0.541 & 0.753 & 1.000 & 0.704 \\
zlib       & 0.496 & 0.663 & 0.994 & 0.639 \\
\bottomrule
\end{tabular}

\caption{Memorization predictor AUROC results for discriminating contamination, for all datasets. Results are shown under the random contamination regime (low, medium, high), with the aggregate binary AUROC shown in the ``all'' column for each benchmark. At low contamination, memorization predictors do not discriminate well but are near perfect for highly contaminated examples. Reference* denotes an oracle method, which uses the standard Hubble model.}
\label{tab:mem_results_full}
\end{table*}
\clearpage

\subsection{Correction results}
\label{app:adjustment_simulation}

\begin{table*}[h]
\centering
\small
\setlength{\tabcolsep}{8pt}
\renewcommand{\arraystretch}{1.2}
\begin{tabular}{lcccccc}
\toprule
\multirow{2}{*}{\textbf{Estimator}} & \multicolumn{3}{c}{\textbf{Random}} & \multicolumn{3}{c}{\textbf{Correlated (high dose)}} \\
\cmidrule(lr){2-4} \cmidrule(lr){5-7}
& \textbf{Low} & \textbf{Mid} & \textbf{High} & \textbf{Easy} & \textbf{Medium} & \textbf{Hard} \\
\midrule
\multicolumn{7}{c}{\textbf{WG (Infill)}} \\
\midrule
Naive & 1.1 & 3.0 & 4.6 & 0.5 & 0.0 & 14.0 \\
IPW (Min-K\%++) & 1.3 & 2.2 & 3.1 & 2.0 & 1.8 & 12.7 \\
Imputation (Llama 3.1) & 1.6 & 1.5 & 2.5 & 2.7 & 3.5 & 11.6 \\
Imputation (Pythia 6.9B) & 1.5 & 1.6 & 1.5 & 3.5 & 5.5 & 7.9 \\
Imputation (Qwen3 8B) & 1.5 & 1.6 & 1.5 & 3.0 & 5.7 & 8.0 \\
Imputation (RoBERTa) & 1.7 & 1.7 & 1.6 & 5.7 & 5.3 & 10.2 \\
Combined (Llama 3.1) & 1.2 & 1.8 & 3.1 & 1.7 & 2.2 & 12.5 \\
Combined (Pythia 6.9B) & 1.1 & 1.3 & 1.7 & 2.3 & 3.6 & 9.9 \\
Combined (Qwen3 8B) & 1.1 & 1.4 & 1.7 & 2.0 & 3.8 & 9.9 \\
Combined (RoBERTa) & 1.2 & 1.2 & 1.7 & 3.9 & 3.6 & 11.4 \\
\midrule
\multicolumn{7}{c}{\textbf{WG (MCQ)}} \\
\midrule
Naive & 1.4 & 1.7 & 3.9 & 4.1 & 6.0 & 1.7 \\
IPW (Min-K\%++) & 1.6 & 1.7 & 2.5 & 4.5 & 5.7 & 4.7 \\
Imputation (Llama 3.1) & 1.8 & 1.6 & 1.6 & 3.8 & 4.8 & 9.6 \\
Imputation (Pythia 6.9B) & 1.7 & 1.6 & 1.7 & 3.0 & 4.6 & 10.8 \\
Imputation (Qwen3 8B) & 1.7 & 1.6 & 1.7 & 3.5 & 4.8 & 11.0 \\
Imputation (RoBERTa) & 1.7 & 1.6 & 1.8 & 4.8 & 4.7 & 12.8 \\
Combined (Llama 3.1) & 1.1 & 1.1 & 1.7 & 3.9 & 5.2 & 6.6 \\
Combined (Pythia 6.9B) & 1.1 & 1.1 & 1.3 & 3.3 & 5.0 & 7.4 \\
Combined (Qwen3 8B) & 1.1 & 1.1 & 1.4 & 3.7 & 5.1 & 7.6 \\
Combined (RoBERTa) & 1.1 & 1.2 & 1.3 & 4.6 & 5.1 & 8.9 \\
\midrule
\multicolumn{7}{c}{\textbf{HellaSwag}} \\
\midrule
Naive & 0.9 & 5.1 & 5.8 & 0.0 & 0.5 & 16.7 \\
IPW (Min-K\%++) & 1.9 & 2.5 & 1.3 & 5.2 & 4.9 & 11.5 \\
Imputation (Llama 3.1) & 1.8 & 5.1 & 6.3 & 1.2 & 1.4 & 17.2 \\
Imputation (Pythia 6.9B) & 1.6 & 1.3 & 1.6 & 1.2 & 3.0 & 6.2 \\
Imputation (Qwen3 8B) & 1.6 & 1.3 & 1.4 & 1.3 & 3.5 & 5.2 \\
Imputation (RoBERTa) & 3.7 & 3.5 & 3.2 & 3.0 & 2.8 & 13.4 \\
Combined (Llama 3.1) & 0.9 & 4.9 & 5.9 & 0.5 & 0.8 & 16.8 \\
Combined (Pythia 6.9B) & 0.9 & 1.7 & 1.5 & 0.9 & 2.8 & 7.1 \\
Combined (Qwen3 8B) & 0.9 & 1.7 & 1.2 & 1.1 & 3.1 & 6.4 \\
Combined (RoBERTa) & 1.1 & 2.9 & 2.1 & 3.7 & 3.4 & 12.4 \\
\bottomrule
\end{tabular}
\caption{Simulated estimator performance on WinoGrande (Infill), WinoGrande (MCQ), and HellaSwag under random and correlated contamination regimes. Values are RMSE against the ground-truth standard model accuracy over 1,000 bootstrap replicates. Estimators using a correctness or memorization predictor reliably adjust for contamination, with the combined estimator using both predictors winning in most settings. Part 1 of 2; see also Table~\ref{tab:all_estimators_part2}.}
\label{tab:all_estimators_part1}
\end{table*}

\begin{table*}[t]
\centering
\small
\setlength{\tabcolsep}{8pt}
\renewcommand{\arraystretch}{1.2}
\begin{tabular}{lcccccc}
\toprule
\multirow{2}{*}{\textbf{Estimator}} & \multicolumn{3}{c}{\textbf{Random}} & \multicolumn{3}{c}{\textbf{Correlated (high dose)}} \\
\cmidrule(lr){2-4} \cmidrule(lr){5-7}
& \textbf{Low} & \textbf{Mid} & \textbf{High} & \textbf{Easy} & \textbf{Medium} & \textbf{Hard} \\
\midrule
\multicolumn{7}{c}{\textbf{MMLU}} \\
\midrule
Naive & 1.8 & 6.4 & 13.1 & 0.5 & 10.3 & 29.2 \\
IPW (Min-K\%++) & 1.7 & 4.5 & 6.4 & 7.5 & 4.1 & 22.3 \\
Imputation (Llama 3.1) & 3.4 & 2.2 & 7.5 & 6.3 & 5.0 & 23.1 \\
Imputation (Pythia 6.9B) & 3.7 & 4.5 & 4.9 & 18.5 & 6.6 & 11.6 \\
Imputation (Qwen3 8B) & 3.5 & 4.3 & 4.1 & 15.5 & 6.1 & 10.5 \\
Imputation (RoBERTa) & 2.7 & 3.2 & 3.6 & 16.8 & 5.5 & 13.2 \\
Combined (Llama 3.1) & 2.3 & 3.5 & 9.6 & 4.0 & 6.9 & 25.4 \\
Combined (Pythia 6.9B) & 2.4 & 1.5 & 1.6 & 13.4 & 2.1 & 16.4 \\
Combined (Qwen3 8B) & 2.3 & 1.4 & 1.8 & 11.2 & 2.0 & 15.4 \\
Combined (RoBERTa) & 1.9 & 1.7 & 2.0 & 12.5 & 1.8 & 17.3 \\
\midrule
\multicolumn{7}{c}{\textbf{PIQA}} \\
\midrule
Naive & 0.9 & 3.0 & 5.3 & 0.0 & 0.0 & 15.9 \\
IPW (Min-K\%++) & 1.7 & 1.4 & 2.5 & 3.1 & 3.0 & 12.9 \\
Imputation (Llama 3.1) & 1.8 & 3.3 & 5.4 & 1.4 & 1.2 & 15.4 \\
Imputation (Pythia 6.9B) & 2.0 & 1.8 & 2.3 & 1.2 & 2.6 & 8.5 \\
Imputation (Qwen3 8B) & 1.5 & 1.6 & 1.6 & 1.6 & 4.0 & 7.7 \\
Imputation (RoBERTa) & 4.0 & 3.4 & 4.6 & 1.7 & 1.8 & 15.0 \\
Combined (Llama 3.1) & 0.9 & 2.7 & 4.9 & 0.6 & 0.8 & 15.1 \\
Combined (Pythia 6.9B) & 0.8 & 1.7 & 2.4 & 1.0 & 2.3 & 9.9 \\
Combined (Qwen3 8B) & 0.9 & 1.5 & 1.9 & 1.3 & 3.2 & 9.4 \\
Combined (RoBERTa) & 1.3 & 2.3 & 3.5 & 2.0 & 2.0 & 14.0 \\
\midrule
\multicolumn{7}{c}{\textbf{PopQA}} \\
\midrule
Naive & 2.4 & 11.6 & 24.0 & 19.2 & 27.9 & 25.0 \\
IPW (Min-K\%++) & 1.3 & 8.9 & 18.2 & 13.2 & 22.2 & 19.4 \\
Imputation (Llama 3.1) & 1.6 & 6.3 & 11.9 & 7.2 & 16.0 & 12.7 \\
Imputation (Pythia 6.9B) & 1.7 & 1.6 & 1.5 & 2.6 & 4.0 & 2.0 \\
Imputation (Qwen3 8B) & 1.5 & 1.6 & 1.4 & 2.3 & 2.5 & 1.3 \\
Imputation (RoBERTa) & 3.7 & 2.9 & 2.9 & 8.1 & 2.0 & 1.7 \\
Combined (Llama 3.1) & 1.6 & 8.2 & 16.3 & 11.5 & 20.4 & 17.2 \\
Combined (Pythia 6.9B) & 1.0 & 4.8 & 9.1 & 6.0 & 12.8 & 8.6 \\
Combined (Qwen3 8B) & 1.1 & 5.0 & 9.1 & 6.2 & 11.9 & 9.2 \\
Combined (RoBERTa) & 1.5 & 3.4 & 7.7 & 2.7 & 11.6 & 9.0 \\
\bottomrule
\end{tabular}
\caption{Simulated estimator performance on MMLU, PIQA, and PopQA under random and correlated contamination regimes. Values are RMSE against the ground-truth standard model accuracy over 1,000 bootstrap replicates. Estimators using a correctness or memorization predictor reliably adjust for contamination, with the combined estimator using both predictors winning in most settings. Part 2 of 2; see also Table~\ref{tab:all_estimators_part1}.}
\label{tab:all_estimators_part2}
\end{table*}
\clearpage

\subsection{EPG estimator}
\begin{table*}[h]
\centering
\small
\setlength{\tabcolsep}{8pt}
\renewcommand{\arraystretch}{1.2}
\begin{tabular}{lcccccc}
\toprule
\multirow{2}{*}{\textbf{Benchmark}} & \multicolumn{3}{c}{\textbf{Random}} & \multicolumn{3}{c}{\textbf{Correlated (high dose)}} \\
\cmidrule(lr){2-4} \cmidrule(lr){5-7}
& \textbf{Low} & \textbf{Mid} & \textbf{High} & \textbf{Easy} & \textbf{Medium} & \textbf{Hard} \\
\midrule
\multicolumn{7}{c}{WG} \\
\midrule
Naive & 1.1 & 3.0 & 4.6 & 0.5 & 0.0 & 14.0 \\
IPW (Min-K\%++) & 1.3 & 2.2 & 3.1 & 2.0 & 1.8 & 12.7 \\
EPG (Min-K\%++)                    & 7.4 & 4.3 & 2.1 & 14.8 & 3.6  & 15.3 
\\
\midrule
\multicolumn{7}{c}{MMLU} \\
\midrule
Naive & 1.8 & 6.4 & 13.1 & 0.5 & 10.3 & 29.2 \\
IPW (Min-K\%++) & 1.7 & 4.5 & 6.4 & 7.5 & 4.1 & 22.3 \\
EPG-IPW (Min-K\%++) & 7.4 & 4.3 & 2.1 & 14.8 & 3.6 & 15.3 \\
\midrule
\multicolumn{7}{c}{PIQA} \\
\midrule
Naive & 0.9 & 3.0 & 5.3 & 0.0 & 0.0 & 15.9 \\
IPW (Min-K\%++) & 1.7 & 1.4 & 2.5 & 3.1 & 3.0 & 12.9 \\
EPG-IPW (Min-K\%++) & 9.5 & 7.7 & 6.5 & 10.8 & 10.7 & 6.6 \\
\midrule
\multicolumn{7}{c}{HellaSwag} \\
\midrule
Naive & 0.9 & 5.1 & 5.8 & 0.0 & 0.5 & 16.7 \\
IPW (Min-K\%++) & 1.9 & 2.5 & 1.3 & 5.2 & 4.9 & 11.5 \\
EPG-IPW (Min-K\%++) & 10.0 & 7.8 & 8.0 & 13.3 & 12.9 & 5.2 \\
\bottomrule
\end{tabular}
\caption{Simulated estimator performance comparing IPW and EPG (spiking-free baseline) using Min-K\%++ as the memorization predictor, across all benchmarks. Values are RMSE against the ground-truth standard model accuracy over 1,000 bootstrap replicates, using a test set of size $n=500$ and contamination rate $\gamma=0.3$. EPG does not use spiking and chooses the contamination threshold heuristically; its gains over the naive estimator are inconsistent across benchmarks, establishing the necessity of spiking.}
\label{tab:epg_all}
\end{table*}
\clearpage

\subsection{Sample efficiency}
\begin{figure}[h]
    \centering
    \includegraphics[width=\columnwidth]{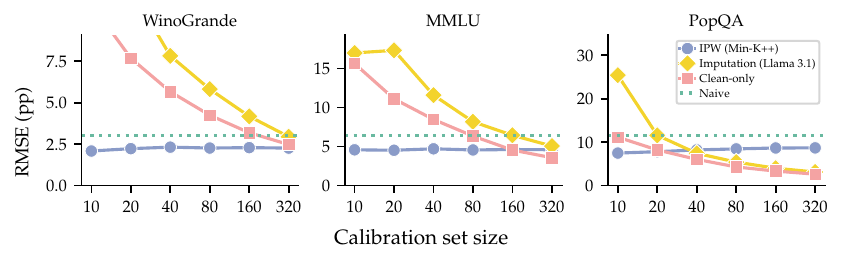}
    \caption{RMSE of IPW (Min-K\%++) and imputation (Llama 3.1) estimators while varying the size of the spiked calibration set, for all datasets. Simulations use mid random contamination and are comparable to those of the same settings in Table~\ref{tab:estimator_random_correlated}. The clean-only estimator uses the calibration items as a new test set and serves as a baseline, while the naive estimator applies no correction. Since items can be spiked at different rates, the calibration set for the memorization predictor is balanced and the IPW estimator is extremely sample efficient.}
    \label{fig:appendix_efficiency}
\end{figure}
\clearpage

\subsection{Dataset transfer}
\label{app:transfer}

\begin{table*}[h]
\centering
\small
\setlength{\tabcolsep}{4pt}
\renewcommand{\arraystretch}{1.1}
\begin{tabular}{llccccc}
\toprule
\textbf{Dose} & \textbf{Source} & \textbf{WG} & \textbf{HellaSwag} & \textbf{MMLU} & \textbf{PIQA} & \textbf{PopQA} \\
\midrule
\multicolumn{7}{c}{\textbf{Low dose}} \\
\midrule
& \textbf{Naive} & 1.1 & 0.9 & 1.8 & 0.9 & 2.4 \\
& \textbf{WG} & 1.4 & 2.7 & 1.7 & 2.6 & 2.8 \\
& \textbf{HellaSwag} & 1.3 & 2.8 & 2.1 & 2.3 & 2.9 \\
& \textbf{MMLU} & 1.2 & 1.9 & 1.5 & 1.8 & 1.7 \\
& \textbf{PIQA} & 1.3 & 2.4 & 1.7 & 2.3 & 2.4 \\
& \textbf{PopQA} & 1.3 & 1.9 & 1.5 & 1.9 & 1.6 \\
& \textbf{Wikipedia} & 1.7 & 5.7 & 3.6 & 4.4 & 5.5 \\
\midrule
\multicolumn{7}{c}{\textbf{Mid dose}} \\
\midrule
& \textbf{Naive} & 3.0 & 5.1 & 6.4 & 3.0 & 11.6 \\
& \textbf{WG} & 1.9 & 1.8 & 5.8 & 1.0 & 2.4 \\
& \textbf{HellaSwag} & 2.0 & 1.4 & 5.7 & 1.0 & 1.7 \\
& \textbf{MMLU} & 2.3 & 2.5 & 5.6 & 1.1 & 4.3 \\
& \textbf{PIQA} & 2.0 & 2.0 & 5.7 & 0.9 & 3.0 \\
& \textbf{PopQA} & 2.2 & 2.9 & 5.8 & 1.1 & 5.0 \\
& \textbf{Wikipedia} & 1.3 & 3.5 & 6.9 & 3.5 & 3.9 \\
\midrule
\multicolumn{7}{c}{\textbf{High dose}} \\
\midrule
& \textbf{Naive} & 4.6 & 5.8 & 13.1 & 5.3 & 24.0 \\
& \textbf{WG} & 2.0 & 1.7 & 5.2 & 1.5 & 3.1 \\
& \textbf{HellaSwag} & 1.8 & 2.4 & 4.0 & 1.8 & 1.1 \\
& \textbf{MMLU} & 2.4 & 1.2 & 4.7 & 1.2 & 5.8 \\
& \textbf{PIQA} & 2.1 & 1.5 & 5.1 & 1.3 & 4.0 \\
& \textbf{PopQA} & 2.6 & 1.4 & 6.7 & 1.4 & 8.6 \\
& \textbf{Wikipedia} & 1.4 & 5.4 & 5.5 & 3.9 & 5.8 \\
\bottomrule
\end{tabular}
\caption{RMSE of the IPW estimator when the memorization predictor (Min-K\%) is transferred across datasets, across low, mid, and high contamination. The memorization predictor is calibrated on the source benchmark and evaluated on all target benchmarks. These simulations are comparable to those of the same settings in Table~\ref{tab:estimator_random_correlated}. At high contamination, transferred predictors almost always improve over the naive estimator, and predictors calibrated on Wikipedia texts closely match dataset-specific calibration.}
\label{tab:transfer_min_k_all_doses}
\end{table*}

\begin{table*}[h]
\centering
\small
\setlength{\tabcolsep}{4pt}
\renewcommand{\arraystretch}{1.1}
\begin{tabular}{llccccc}
\toprule
\textbf{Dose} & \textbf{Source} & \textbf{WG} & \textbf{HellaSwag} & \textbf{MMLU} & \textbf{PIQA} & \textbf{PopQA} \\
\midrule
\multicolumn{7}{c}{\textbf{Low dose}} \\
\midrule
& \textbf{Naive} & 1.1 & 0.9 & 1.8 & 0.9 & 2.4 \\
& \textbf{WG} & 1.3 & 1.3 & 1.8 & 1.6 & 1.0 \\
& \textbf{HellaSwag} & 1.5 & 1.9 & 1.8 & 2.0 & 2.1 \\
& \textbf{MMLU} & 1.4 & 1.6 & 1.7 & 1.7 & 1.4 \\
& \textbf{PIQA} & 1.4 & 1.4 & 1.8 & 1.7 & 1.0 \\
& \textbf{PopQA} & 1.2 & 1.1 & 1.8 & 1.3 & 1.3 \\
& \textbf{Wikipedia} & 1.1 & 0.9 & 1.8 & 0.9 & 2.4 \\
\midrule
\multicolumn{7}{c}{\textbf{Mid dose}} \\
\midrule
& \textbf{Naive} & 3.0 & 5.1 & 6.4 & 3.0 & 11.6 \\
& \textbf{WG} & 2.2 & 3.7 & 5.2 & 1.6 & 7.1 \\
& \textbf{HellaSwag} & 1.7 & 2.5 & 4.2 & 1.0 & 3.1 \\
& \textbf{MMLU} & 2.0 & 3.0 & 4.5 & 1.2 & 4.7 \\
& \textbf{PIQA} & 2.1 & 3.5 & 5.1 & 1.4 & 6.5 \\
& \textbf{PopQA} & 2.5 & 4.3 & 5.7 & 2.1 & 8.9 \\
& \textbf{Wikipedia} & 3.0 & 5.1 & 6.4 & 3.0 & 11.6 \\
\midrule
\multicolumn{7}{c}{\textbf{High dose}} \\
\midrule
& \textbf{Naive} & 4.6 & 5.8 & 13.1 & 5.3 & 24.0 \\
& \textbf{WG} & 3.1 & 3.2 & 9.1 & 2.8 & 14.4 \\
& \textbf{HellaSwag} & 2.1 & 1.3 & 5.3 & 1.3 & 5.7 \\
& \textbf{MMLU} & 2.4 & 1.8 & 6.4 & 1.7 & 8.8 \\
& \textbf{PIQA} & 3.0 & 2.9 & 8.6 & 2.5 & 13.2 \\
& \textbf{PopQA} & 3.7 & 4.2 & 10.7 & 3.7 & 18.2 \\
& \textbf{Wikipedia} & 4.6 & 5.8 & 13.1 & 5.3 & 24.0 \\
\bottomrule
\end{tabular}
\caption{RMSE of the IPW estimator when the memorization predictor (Min-K\%++) is transferred across datasets, across low, mid, and high contamination. The memorization predictor is calibrated on the source benchmark and evaluated on all target benchmarks. These simulations are comparable to those of the same settings in Table~\ref{tab:estimator_random_correlated}. Min-K\%++ is less transferable than Min-K\% due to dataset-specific normalization, though at high contamination transferred predictors still generally improve over the naive estimator.}
\label{tab:transfer_min_k_plus_plus_all_doses}
\end{table*}

\clearpage

\end{document}